\documentclass[12pt]{article}
\newtheorem{theorem}{Theorem}

\input epsf.tex

\begin{document}

\begin{center}
{\Large \bf  Robust designs for experiments with blocks}

Rena K. Mann, Roderick Edwards  and Julie Zhou\footnote{
Corresponding author;  Phone: (250) 721-7470,  Email:  jzhou@uvic.ca}

Department of Mathematics and Statistics \\

University of Victoria, Victoria, BC, Canada V8W 2Y2 \\

\end{center}



\bigskip

\noindent
ABSTRACT

{\small
For experiments running in field plots or over time, the observations  are 
often correlated due to spatial or serial correlation, 
which  leads to correlated errors  in a linear model 
analyzing the treatment means. 
Without knowing the exact correlation matrix of the errors, it is not possible to compute the generalized
least squares estimator for the treatment means and use it to construct optimal designs for the experiments.
In this paper we propose to use neighbourhoods to model the covariance matrix of the errors, and apply a
modified generalized least squares estimator to construct robust designs for experiments with blocks.
A minimax design criterion is investigated,  and 
a simulated annealing algorithm is developed to find robust designs.  We have derived  several theoretical results, and
representative examples are presented.  
}

\section{Introduction}

Consider a linear regression model,
\begin{eqnarray}
y_i={\bf z}^\top({\bf x}_i) \mbox{\boldmath $\theta$} + \epsilon_i, ~~~i=1, \cdots, N,
\label{eqn1.1}
\end{eqnarray}
where the response variable $y_i$ is observed at design point ${\bf x}_i$ from design space $S \in R^q$, 
${\bf z}({\bf x})$ is a  known function of ${\bf x}$,
parameter vector $\mbox{\boldmath $\theta$}$ belongs to $R^p$, 
and the errors $\epsilon_i$ are uncorrelated and have mean zero and variance $\sigma^2$.
Let $\mbox{\boldmath $\epsilon$}=(\epsilon_1, \cdots, \epsilon_N)^\top$, then $Cov(\mbox{\boldmath $\epsilon$})=\sigma^2 {\bf I}_N$,
where ${\bf I}_N$ is the $N \times N $ identity matrix.
The least squares estimator (LSE) of $\mbox{\boldmath $\theta$}$ is given by
$\hat{\mbox{\boldmath $\theta$}}=\left({\bf Z}^\top {\bf Z}\right)^{-1} {\bf Z}^\top {\bf y}$, 
where ${\bf Z}=({\bf z}({\bf x}_1), \cdots, {\bf z}({\bf x}_N))^\top$ is the model matrix and 
${\bf y}=(y_1, \cdots, y_N)^\top$ is the vector of responses.  
The LSE is the best linear unbiased estimator (BLUE) and its covariance matrix is 
$Cov(\hat{\mbox{\boldmath $\theta$}}) = \sigma^2 \left({\bf Z}^\top {\bf Z}\right)^{-1}$.

Optimal regression designs, minimizing some scalar functions of $Cov(\hat{\mbox{\boldmath $\theta$}})$ 
over the choices of ${\bf x}_1, \cdots, {\bf x}_N \in S$,  have been investigated extensively in the literature, 
for example,  see Fedorov (2010) and Pukelsheim (1993).
Commonly used design criteria include D-optimal and A-optimal criteria.  
D-optimal and A-optimal designs minimize, respectively, 
the determinant and the trace  of $Cov(\hat{\mbox{\boldmath $\theta$}})$.  
However, these optimal designs are very sensitive to the model assumptions.
If there are violations to the model assumptions, the optimal designs may produce large variance and/or large bias of 
$\hat{\mbox{\boldmath $\theta$}}$.  
Therefore robust regression designs have been investigated against various departures from the model.

Robust designs against the misspecification of the response surface function were  studied by, 
among many others,  Box and Draper (1959),  Huber (1975), and  Wiens (1992).  
Robust designs against the autocorrelation among the errors were explored in, for example, Bickel and Herzberg (1979) and
Bickel, Herzberg and Schilling (1981).  
Developments of robust designs against both departures in the response surface function and in the autocorrelation
include Wiens and Zhou (1997, 1999), Shi, Ye and Zhou (2007), and Zhou (2001).

In the design of experiments, due to randomization of experimental runs, the errors in the linear models
investigating the effects of factors are usually considered to be uncorrelated. 
See, for example, Montgomery (2012).
However, for some experiments including field experiments, 
the errors are often correlated,  which has been recognized by many researchers, such as
Williams (1952), Herzberg (1982), Martin (1982, 1986).
In particular, serial correlation over time (or the order of experimental runs) and spatial correlation over field plots are quite common.
Since regression models can be built to analyze the factor effects, optimal or robust designs of experiments can be studied similarly as in
optimal or robust regression designs.  
This leads to the research in Wiens and Zhou (2008) and Ou and Zhou (2009) using the minimax approach to find 
robust designs for field experiments.  
The designs in Ou and Zhou (2009) are robust against departures from the covariance
structure of the errors,   while the designs in Wiens and Zhou (2008) are robust against departures from both the covariance 
structure of the errors and the response function. 

In this paper, we extend the work in Wiens and Zhou (2008) and Ou and Zhou (2009) to find robust designs for experiments
that need to be performed in several blocks. Because of the blocks, 
the covariance structure of the errors is more complicated than those in
Wiens and Zhou (2008) and Ou and Zhou (2009).  The  designs we construct here are robust against departures from
the covariance structure of the errors.  The applications include experiments with serial correlation or spatial 
correlation.

The rest of the paper is organized as follows. 
In Section 2, we first discuss the linear model to analyze experiments with blocks and its related
regression model, and present the least squares estimator and the generalized least squares estimator. 
Then we use one experiment with blocks to illustrate the influence of the error correlation 
on the covariance of the LSE. 
In Section 3, two neighborhoods of the covariance matrix of the errors are defined. 
Based on the covariance neighborhoods, a robust 
design criterion is proposed. 
In Section 4, robust designs are studied and constructed.  
A simulated annealing algorithm is  developed to compute robust designs, and 
applications are presented. In addition,  several theoretical results are derived.
Concluding remarks are in Section 5.  
All proofs are given in the Appendix.

\section{Linear models and estimators}

\subsection{Linear models}

Consider experiments with blocks to compare $t$ treatment means,
where blocking is  used to eliminate nuisance sources of variability in the experiments.
Complete block designs allow one replicate of $t$ treatment runs within each block,
while incomplete block designs have less than $t$ runs in each block.  
In this paper we consider complete block designs and assume the block effects are fixed.  
Suppose there are $b$ blocks with $b \ge 2$, and treatments are  numbered as $1, 2, \cdots, t$. 
The linear effects model can be written as
\begin{eqnarray}
y_{ij}=\mu+\tau_r+\beta_j+\epsilon_{ij}, ~~~~i=1, \cdots, t, ~j=1, \cdots, b,
\label{eqn2.1}
\end{eqnarray}
where $y_{ij}$ is the $i$th response  in the $j$th block from the $r$th treatment ($r=1, \cdots, t$), $\mu$ is the overall mean,
$\tau_r$ is the $r$th treatment effect, 
$\beta_j$ is the $j$th block effect, and
$\epsilon_{ij}$ is the random error term.  In order to identify the parameters uniquely in the model, 
the treatment effects and block effects satisfy constraints
$\sum_{r=1}^t \tau_r=0$ and $\sum_{j=1}^b \beta_j =0$.

A regression model can also be used to analyze the treatment effects.  
Define the following vectors and matrices to
present the regression model:
\begin{eqnarray*}
{\bf y}_j=\left( \begin{array}{c}
y_{1j} \\
\vdots \\
y_{tj} 
\end{array} \right),
~{\bf y}=\left( \begin{array}{c}
{\bf y}_{1} \\
\vdots \\
{\bf y}_{b} 
\end{array} \right),
~\mbox{\boldmath $\mu$}=\left( \begin{array}{c}
\mu_1 \\
\vdots \\
\mu_t 
\end{array} \right)=\left( \begin{array}{c}
\mu+\tau_1 \\
\vdots \\
\mu+\tau_t 
\end{array} \right),
~\mbox{\boldmath $\beta$}
=\left( \begin{array}{c}
\beta_1 \\
\vdots \\
\beta_{b-1}
\end{array} \right),
\end{eqnarray*}
and
\begin{eqnarray*}
\mbox{\boldmath $\epsilon$}_j=\left( \begin{array}{c}
\epsilon_{1j} \\
\vdots \\
\epsilon_{tj} 
\end{array} \right),
~\mbox{\boldmath $\epsilon$}=\left( \begin{array}{c}
\mbox{\boldmath $\epsilon$}_1 \\
\vdots \\
\mbox{\boldmath $\epsilon$}_b 
\end{array} \right),
~~{\bf X}=\left( \begin{array}{c}
 {\bf X}_1\\
\vdots  \\
{\bf X}_b
\end{array} \right),
~~~{\bf U}=\left( \begin{array}{c}
 {\bf U}_1\\
\vdots  \\
{\bf U}_b
\end{array} \right),
\end{eqnarray*}
where matrix ${\bf X}_j$ ($t \times t$) is the design matrix for the $t$ treatments in block $j$,  $j=1, \cdots, b$, and 
matrix ${\bf U}$ is the model matrix for the block effects. 
The elements of ${\bf X}_j$ are either $0$ or $1$,  and each row has only one $1$.  If $y_{ij}$ received treatment $r$, 
then the element at the $i$th row and $r$th column of
${\bf X}_j$ is 1. Since $\sum_{j=1}^b \beta_j =0$, parameter $\beta_b=-\beta_1 - \cdots - \beta_{b-1}$.
Therefore we only need $b-1$ parameters in vector $\mbox{\boldmath $\beta$} $  for the regression model,
and matrices ${\bf U}_1, \cdots, {\bf U}_b$ ($t \times (b-1)$) for the $b$ blocks are given by
\begin{eqnarray*}
{\bf U}_1=\left( \begin{array}{cccc}
1 & 0 & \cdots & 0 \\
\vdots & \vdots & \vdots & \vdots  \\
1 & 0 & \cdots & 0 
\end{array} \right),
\cdots,
{\bf U}_{b-1}=\left( \begin{array}{cccc}
0 &  \cdots & 0& 1\\
\vdots & \vdots & \vdots & \vdots  \\
0&  \cdots & 0 & 1
\end{array} \right),
{\bf U}_b=\left( \begin{array}{ccc}
-1 &  \cdots & -1 \\
\vdots &  \vdots & \vdots  \\
-1 &  \cdots & -1
\end{array} \right).
\end{eqnarray*}
Now the regression model for the effects model (\ref{eqn2.1}) is given by
\begin{eqnarray}
{\bf y}={\bf X} \mbox{\boldmath $\mu$} + {\bf U} \mbox{\boldmath $\beta$} + \mbox{\boldmath $\epsilon$}.
\label{reg1}
\end{eqnarray}
Notice that there is no grand mean (or intercept) in this model, 
since vector $\mbox{\boldmath $\mu$}$ includes the grand mean component $\mu$ in each $\mu_r$, $r=1, \cdots, t$.

\subsection{Estimators}

In order to estimate $\mbox{\boldmath $\mu$}$ and $\mbox{\boldmath $\beta$}$ efficiently, it is important to
know the covariance matrix of the error vector $\mbox{\boldmath $\epsilon$}$.
Two cases are discussed below.

\begin{description}
\item{Case (i):}  The errors are uncorrelated, {\it i.e.}, $Cov(\mbox{\boldmath $\epsilon$})=\sigma^2 {\bf I}_N$, where $N=t b$.

\item{Case (ii):}  The errors are correlated, {\it i.e.}, $Cov(\mbox{\boldmath $\epsilon$})=\sigma^2 {\bf V}$, where ${\bf V}$ is an
$N \times N$ correlation matrix. 
In particular, there may be correlation among the errors within each block.  This includes the situations in which the runs in each block
are conducted over time or the runs are located in field plots. Assume the errors between blocks are independent. 
Let ${\bf V}_j$ be the correlation matrix for the errors in block
$j$, $j=1, \cdots, b$, so ${\bf V}$ is a block diagonal matrix, {\it i.e.}, 
${\bf V}= {\bf V}_1 \oplus{\bf V}_2 \oplus \cdots \oplus {\bf V}_b$.
\end{description}

Define
\begin{eqnarray*}
\mbox{\boldmath $\theta$} = \left( \begin{array}{c}
\mbox{\boldmath $\mu$} \\
\mbox{\boldmath $\beta$} 
\end{array} \right), ~~~
{\bf Z}=\left({\bf X}, {\bf U} \right).
\end{eqnarray*}
Then model (\ref{reg1}) becomes, ${\bf y}={\bf Z}\mbox{\boldmath $\theta$} + \mbox{\boldmath $\epsilon$}$.
The LSE and the generalized least squares estimator (GLSE) are, respectively,
\begin{eqnarray}
&&\hat{\mbox{\boldmath $\theta$}}_{L}=\left( \begin{array}{c}
\hat{\mbox{\boldmath $\mu$}}_{L} \\
\hat{\mbox{\boldmath $\beta$}}_{L}
\end{array} \right)
=\left({\bf Z}^\top {\bf Z}\right)^{-1} {\bf Z}^\top {\bf y},
\label{LSE1} \\
&&\hat{\mbox{\boldmath $\theta$}}_{G}=\left( \begin{array}{c}
\hat{\mbox{\boldmath $\mu$}}_{G} \\
\hat{\mbox{\boldmath $\beta$}}_{G}
\end{array} \right)
=\left({\bf Z}^\top {\bf V}^{-1} {\bf Z}\right)^{-1} {\bf Z}^\top {\bf V}^{-1} {\bf y}.
\label{GLSE1} 
\end{eqnarray}

From Section 2.1, it is easy to verify that 
\begin{eqnarray*}
{\bf Z}^\top {\bf Z}=\left(
\begin{array}{cc}
{\bf X}^\top {\bf X} & {\bf X}^\top {\bf U} \\
{\bf U}^\top {\bf X} & {\bf U}^\top {\bf U}
\end{array} \right)
=\left(
\begin{array}{cc}
b ~{\bf I}_t  & {\bf 0} \\
{\bf 0}  & {\bf U}^\top {\bf U}
\end{array} \right),
\end{eqnarray*}
which implies that the regressors for $\mbox{\boldmath $\mu$}$ and $\mbox{\boldmath $\beta$} $ are orthogonal.
For block designs, we are mainly interested in estimating and comparing the treatment effects, so we will look at the variances of 
$\hat{\mbox{\boldmath $\mu$}}_{L} $ and
$\hat{\mbox{\boldmath $\mu$}}_{G}$ to construct optimal/robust designs in Sections 3 and 4.

For Case (i), the LSE is the BLUE, and
\begin{eqnarray}
Cov \left(\hat{\mbox{\boldmath $\mu$}}_{L}\right) &=& \sigma^2 \left(  {\bf X}^\top {\bf X} \right)^{-1} 
= \frac{\sigma^2}{b} ~{\bf I}_t. 
\label{CovLSE1} 
\end{eqnarray}

For Case (ii), the GLSE is the BLUE, and
\begin{eqnarray}
Cov \left(\hat{\mbox{\boldmath $\mu$}}_{L}\right) &=& \sigma^2  \left(  {\bf X}^\top {\bf X} \right)^{-1}
{\bf X}^\top {\bf V} {\bf X}  \left(  {\bf X}^\top {\bf X} \right)^{-1}
=\frac{\sigma^2}{b^2} ~{\bf X}^\top {\bf V} {\bf X},
\label{CovLSE2} \\
Cov \left(\hat{\mbox{\boldmath $\mu$}}_{G}\right) &=& \sigma^2  {\bf C}_{\mu},  
\label{CovGLSE2}
\end{eqnarray}
where matrix ${\bf C}_{\mu}$ is the submatrix of $\left({\bf Z}^\top {\bf V}^{-1} {\bf Z}\right)^{-1}$, consisting of the
first $t$ rows and the first $t$ columns.

\subsection{An example}

We use one example of randomized complete block design to illustrate the influence of 
the error correlation on the covariance of the LSE in (\ref{CovLSE2}).

There is one example of a randomized complete block design in Montgomery (2012, page 178) to study the effect
of three different lubricating oils (treatments) on fuel consumption in diesel truck engines.   
Five different truck engines are available for the experiment.
Since there may be differences among truck engines, a randomized complete block design is used, where the five
truck engines are the five blocks.  The observed data on fuel consumption are given in Table \ref{table1}.

\begin{table}
\caption{Fuel consumption data}
\begin{center}
\begin{tabular}{cccccc}  \hline
 & & & Truck & & \\
 Oil & 1 & 2 & 3 & 4 & 5 \\  \hline
1 & 0.500 & 0.634 & 0.487 & 0.329 & 0.512 \\
2 & 0.535 & 0.675 & 0.520 & 0.435 & 0.540 \\
3 & 0.513 & 0.595 & 0.488 & 0.400 & 0.510 \\ \hline
\end{tabular}
\end{center}
\label{table1}
\end{table}
 
We use model (\ref{reg1}) to analyze the treatment means, where 
$\mbox{\boldmath $\mu$}=(\mu_1, \mu_2, \mu_3)^\top$,  and
$\mbox{\boldmath $\beta$}=(\beta_1,\beta_2,\beta_3,\beta_4)^\top$.
Since we do not know the run order in each block, we just use the standard order in Table {\ref{table1}.
So the design matrices are 
\begin{eqnarray}
{\bf X}_j=\left( \begin{array}{ccc}
1  & 0  & 0 \\
0  & 1  & 0 \\
0  & 0  & 1
\end{array} \right), ~~~j=1, \cdots, 5.
\label{Xj}
\end{eqnarray}

Using the LSE, we get  
$\hat{\mbox{\boldmath $\mu$}}_{L}=(0.492, 0.541, 0.501)^\top$ and
$\hat{\sigma}=0.023$.
If the errors are uncorrelated, then from (\ref{CovLSE1}) the covariance matrix of $\hat{\mbox{\boldmath $\mu$}}_{L}$
is 
\begin{eqnarray*}
Cov \left(\hat{\mbox{\boldmath $\mu$}}_{L}\right) = \sigma^2 \left(  {\bf X}^\top {\bf X} \right)^{-1} 
= \frac{\sigma^2}{5} ~{\bf I}_3. 
\end{eqnarray*}
The run order does not affect this covariance when the errors are uncorrelated.
Thus the estimated standard error for $\hat{\mu}_i$ is $\hat{\sigma}/\sqrt{5}=0.010$, for all $i=1, 2, 3$. Inferences can be made for 
any linear functions of $\mbox{\boldmath $\mu$}$.

If the errors are correlated, then from (\ref{CovLSE2}) the covariance matrix of $\hat{\mbox{\boldmath $\mu$}}_{L}$
is 
\begin{eqnarray}
Cov \left(\hat{\mbox{\boldmath $\mu$}}_{L}\right) 
=\frac{\sigma^2}{5^2} ~{\bf X}^\top  {\bf V} {\bf X}. 
\label{CovLSE3}
\end{eqnarray}
In this case, the run order affects  $Cov \left(\hat{\mbox{\boldmath $\mu$}}_{L}\right)$.
Suppose ${\bf V}= {\bf V}_1 \oplus {\bf V}_2 \oplus \cdots \oplus {\bf V}_5$.  Since the runs 
in this experiment are conducted over time, it may be reasonable to model ${\bf V}_j$ with a nearest neighbour
correlation matrix with 
\begin{eqnarray*}
{\bf V}_j = \left( \begin{array}{ccc}
1 & \rho & 0 \\
\rho & 1 & \rho \\
0  & \rho & 1 
\end{array} \right),  ~~~~\rho \in [-0.25, 0.25], ~~j=1, \cdots, 5.
\end{eqnarray*}

Define ${\bf A}(d) = {\bf X}^\top  {\bf V} {\bf X}$, where $d$ indicates a design.  Notice that design matrices
${\bf X}_1, \cdots, {\bf X}_b$ depend on the run order in each block, and each can be obtained by permuting the rows 
of the matrix in (\ref{Xj}).
Then the covariance matrix in (\ref{CovLSE3}) is affected by the run order of the three treatments in each block.
Consider the three designs in Table \ref{table2} and $\rho=0.2$.  
It can be easily shown that,
\begin{eqnarray*}
&&  {\bf A}(d1) =\left(
\begin{array}{ccc}
5 & 1 & 0 \\
1 & 5 & 1 \\
0 & 1 & 5 
\end{array} \right), ~~~
 {\bf A}(d2)=\left(
\begin{array}{ccc}
5.0 & 0.8 & 0.6 \\
0.8 & 5.0 & 0.6 \\
0.6 & 0.6 & 5.0 
\end{array} \right), ~~~\mbox{and} \\
&&  {\bf A}(d3) =\left(
\begin{array}{ccc}
5.0 & 0.8 & 0.8 \\
0.8 & 5.0 & 0.4 \\
0.8 & 0.4 & 5.0 
\end{array} \right),  
\end{eqnarray*}
and  $\det\left({\bf A}(d1)\right)=115, 
~\det\left({\bf A}(d2)\right)=118.776$, $\det\left({\bf A}(d3)\right)=118.312$.
It is clear that the run order affects the $Cov \left(\hat{\mbox{\boldmath $\mu$}}_{L}\right)$ in (\ref{CovLSE2}).
Similarly we can show that the run order also affects the $Cov \left(\hat{\mbox{\boldmath $\mu$}}_{G}\right)$ in (\ref{CovGLSE2}).

\begin{table}
\caption{Three block designs}
\begin{center}
\begin{tabular}{c|ccccc}  \hline
Design/run order & block 1 & block 2 & block 3 & block 4 & block 5 \\
 \hline
d1 & 1, 2, 3 & 3, 2, 1& 3, 2, 1 & 1, 2, 3  & 3, 2, 1  \\ \hline
d2 & 1, 2, 3 & 2, 1, 3 & 1, 3, 2 & 3, 2, 1 & 3, 1, 2 \\ \hline
d3 & 3, 2, 1 & 2, 1, 3  & 2, 1, 3 & 2, 1, 3 & 1, 3, 2\\ \hline
\end{tabular}
\end{center}
\label{table2}
\end{table}

In practice, if we do not have any information on the correlation matrix ${\bf V}$, the randomized run order should be
used in each block.
However, if we have some information on the correlation matrix ${\bf V}$, we can use an ``optimal" run order in each
block to minimize the $Cov \left(\hat{\mbox{\boldmath $\mu$}}_{L}\right)$ or 
$Cov \left(\hat{\mbox{\boldmath $\mu$}}_{G}\right)$.  In the next Section, we will propose a robust design criterion to 
find the ``optimal" run order.

%
%
%
%
%
%

\section{Minimax design criterion}

For practical applications, we never know the exact covariance of the errors in model (\ref{reg1}), but we may have some information
about the correlation structure.   A flexible model for the 
$Cov(\mbox{\boldmath $\epsilon$})$ is to use a neighborhood of covariance matrices, which is defined in Section 3.1.
Since we do not know the ${\bf V}$ in the GLSE in (\ref{GLSE1}), we will modify it in Section 3.2 using the information in 
the neighborhood of $Cov(\mbox{\boldmath $\epsilon$})$. Based on the neighborhood of $Cov(\mbox{\boldmath $\epsilon$})$
and the modified GLSE, a robust design criterion is given in Section 3.3 to construct the optimal run order in each block.

\subsection{Neighbourhoods of covariance matrices}

Let ${\bf R}=\sigma^2 {\bf V}=\sigma^2{\bf V}_1 \oplus \sigma^2{\bf V}_2 \oplus \cdots \oplus \sigma^2{\bf V}_b$.
In Mann (2011), two neighbourhoods of ${\bf R}$ were proposed, which are extensions of the 
neighbourhoods of covariance matrices in Wiens and Zhou (2008). We briefly describe them below. 

Suppose ${\bf R}_0=\sigma^2 {\bf V}_0=\sigma^2{\bf V}_{10} \oplus \sigma^2{\bf V}_{20} \oplus \cdots \oplus \sigma^2{\bf V}_{b0}$,
where 
${\bf V}_{10}, \cdots, {\bf V}_{b0}$ are known correlation matrices.  Often
${\bf V}_{10}, \cdots, {\bf V}_{b0}$ are viewed as our prior knowledge of the error process in model (\ref{reg1}).
Commonly used error processes for field plots include the nearest neighbour (NN), moving average (MA), doubly geometric (DG)
and discrete exponential (DE) processes, which are reviewed in detail in Mann (2011). Two options for
neighbourhoods of ${\bf R}$ are defined
around ${\bf R}_0$ using the following procedure.

\begin{description}
\item{(i)} Let ${\bf R}_{j0}=\sigma^2{\bf V}_{j0}$, for $j=1, \cdots,b$.

\item{(ii)} Define a class of covariance matrices around each ${\bf R}_{j0}$, $j=1, \cdots, b$,
\begin{eqnarray*}
{\bf R}_{j, {\bf K}_j, \alpha}=\left\{ ~{\bf B} ~|~ 0 \le {\bf B} \le {\bf R}_{j0} + \alpha {\bf K}_j, ~{\bf B}^\top={\bf B} ~\right\},
\end{eqnarray*}
where $\alpha \ge 0$, and ${\bf K}_j$ is either ${\bf R}_{j0}$ or ${\bf I}_t$. The matrix ordering is by positive semi-definiteness,
{\it i.e.}, ${\bf B} \ge 0$ means that ${\bf B}$ is positive semi-definite.
For the applications in Section 4.2, we take ${\bf R}_{10}={\bf R}_{20}=\cdots={\bf R}_{b0}$, so ${\bf K}_j$ does not depend on $j$.
Thus, for simplicity, we omit the subscript $j$ in ${\bf K}_j$ in the following.

\item{(iii)} Define two neighbourhoods of ${\bf R}$,
\begin{eqnarray}
{\bf R}_{{\bf K}, \alpha}=\left\{ ~{\bf R} ~|~ {\bf R}={\bf B}_1 \oplus {\bf B}_2 \oplus \cdots  \oplus {\bf B}_b, ~
{\bf B}_j \in {\bf R}_{j, {\bf K}, \alpha}, ~j=1, \cdots, b ~\right\},
\label{NBR}
\end{eqnarray}
where ${\bf K}$ is either ${\bf R}_{j0}$ or ${\bf I}_t$.  So
the two neighbourhoods are ${\bf R}_{{\bf R}_{j0}, \alpha}$ and ${\bf R}_{{\bf I}_t, \alpha}$.

\end{description}

We can also use matrix norms $|| \cdot ||_1$ or $|| \cdot ||_2$ 
(Horn and Johnson, 1985, page 291) to define a neighbourhood of ${\bf R}$. Let
\begin{eqnarray*}
{\bf R}_{l, \alpha}=\left\{ ~{\bf R} ~|~ {\bf R}={\bf B}_1 \oplus {\bf B}_2 \oplus \cdots  \oplus {\bf B}_b, ~
||{\bf B}_j - {\bf R}_{j0}||_l \le \alpha, ~{\bf B}_j^\top = {\bf B}_j \ge 0, ~j=1, \cdots, b ~\right\},
\end{eqnarray*}
$l=1, ~2$.  
However, it is shown in Wiens and Zhou (2008) that 
${\bf R}_{l, \alpha} = {\bf R}_{{\bf K}, \alpha}$ with ${\bf K}={\bf I}_t$.  Thus we will only use 
the neighbourhoods ${\bf R}_{{\bf K}, \alpha}$ to define and construct robust designs in this paper.

It is obvious that parameter $\alpha$ controls the size of the neighbourhoods of ${\bf R}$. The larger the $\alpha$ is,
the bigger the neighbourhood is.  It is also clear that 
${\bf R}_0  \in {\bf R}_{{\bf K}, \alpha}$ for all $\alpha \ge 0$, and ${\bf R}_0$ can be viewed as a center of the
neighbourhoods.

\subsection{Modified GLSE}

We cannot compute the  GLSE in (\ref{GLSE1}) without knowing matrix ${\bf V}$. A modified GLSE (MGLSE) is proposed 
when $ {\bf R}=Cov(\mbox{\boldmath $\epsilon$})$ belongs to 
${\bf R}_{{\bf K}, \alpha}$. 
The original idea is from Martin (1986), but it is applied for ${\bf R}_{{\bf K}, \alpha}$ in Mann (2011). 
Define the MGLSE as
\begin{eqnarray}
\hat{\mbox{\boldmath $\theta$}}_{M}=\left( \begin{array}{c}
\hat{\mbox{\boldmath $\mu$}}_{M} \\
\hat{\mbox{\boldmath $\beta$}}_{M}
\end{array} \right)
=\left({\bf Z}^\top {\bf R}_0^{-1} {\bf Z}\right)^{-1} {\bf Z}^\top {\bf R}_0^{-1} {\bf y}.
\label{GLSE2} 
\end{eqnarray}
Then the covariance matrix of $\hat{\mbox{\boldmath $\mu$}}_{M}$ is
\begin{eqnarray}
Cov\left( \hat{\mbox{\boldmath $\mu$}}_{M} \right) = {\bf T} 
\left({\bf Z}^\top {\bf R}_0^{-1} {\bf Z}\right)^{-1}
{\bf Z}^\top {\bf R}_0^{-1}  {\bf R}  {\bf R}_0^{-1} {\bf Z}
\left({\bf Z}^\top {\bf R}_0^{-1} {\bf Z}\right)^{-1} 
{\bf T}^\top,
\label{COVM}
\end{eqnarray}
where ${\bf T}=\left( {\bf I}_t, {\bf 0} \right)$ is a  $t \times (t+b-1)$ matrix, and 
${\bf R}$ is the true (but unknown) covariance matrix of the errors. 

\subsection{Design criterion}

Suppose $\hat{\mbox{\boldmath $\mu$}}$ is an estimator of ${\mbox{\boldmath $\mu$}}$,
such as the LSE or the MGLSE.
Let function $g_{\cal L} \left( \hat{\mbox{\boldmath $\mu$}}, {\bf X},  {\bf R} \right) 
= {\cal L} \left(  Cov\left( \hat{\mbox{\boldmath $\mu$}}\right) \right)$ be a measure of the
covariance matrix.  The commonly used measures ${\cal L}$ include the determinant and trace.
Function $g_{\cal L}$ depends on the estimator $\hat{\mbox{\boldmath $\mu$}}$, model matrix ${\bf Z}$ and 
the error covariance matrix ${\bf R}$;
see (\ref{CovLSE2}) and (\ref{COVM}).
Since matrix ${\bf U}$ is fixed in ${\bf Z}$, we write $g_{\cal L}$ depending  on ${\bf Z}$ only through ${\bf X}$.

Since the covariance matrix of $\hat{\mbox{\boldmath $\mu$}}$ depends on the unknown ${\bf R}$, 
we cannot minimize
$g_{\cal L} \left( \hat{\mbox{\boldmath $\mu$}}, {\bf X},  {\bf R} \right) $
directly to construct optimal designs.
A minimax approach has been used to construct robust designs for various models. See, for example,
Huber (1975), Wiens (1992), and Ou and Zhou (2009).
The minimax approach will also be  applied here to define robust designs.

Define the maximum loss function as 
\begin{eqnarray}
\phi_{\cal L} 
\left( \hat{\mbox{\boldmath $\mu$}}, {\bf X} \right) 
=\max_{{\bf R} \in {\bf R}_{{\bf K}, \alpha}} g_{\cal L} \left( \hat{\mbox{\boldmath $\mu$}}, {\bf X},  {\bf R} \right). 
\label{maxLoss1}
\end{eqnarray}
Use $\phi_{A}$ or $\phi_{D}$ to denote the $\phi_{\cal L}$ when measure ${\cal L}$ is the trace or determinant respectively.
A minimax (robust) design $\xi_{\cal L} $ is defined to be the design that minimizes
$\phi_{\cal L} 
\left( \hat{\mbox{\boldmath $\mu$}}, {\bf X} \right) $ over design matrix ${\bf X}$.

From the definition,
the minimax  design may depend on the estimator $\hat{\mbox{\boldmath $\mu$}}$.
For the LSE, from (\ref{CovLSE2}) and ${\bf R}=\sigma^2{\bf V}$,
\begin{eqnarray}
g_{\cal L} \left( \hat{\mbox{\boldmath $\mu$}}_L, {\bf X},  {\bf R} \right)
={\cal L} \left(  Cov\left( \hat{\mbox{\boldmath $\mu$}}_L\right) \right)
={\cal L} \left( \frac{1}{b^2} ~{\bf X}^\top {\bf R} {\bf X} \right).
\label{gloss1}
\end{eqnarray}
For the MGLSE, from (\ref{COVM}),
\begin{eqnarray}
g_{\cal L} \left( \hat{\mbox{\boldmath $\mu$}}_M, {\bf X},  {\bf R} \right)
={\cal L} \left(
{\bf T} 
\left({\bf Z}^\top {\bf R}_0^{-1} {\bf Z}\right)^{-1}
{\bf Z}^\top {\bf R}_0^{-1}  {\bf R}  {\bf R}_0^{-1} {\bf Z}
\left({\bf Z}^\top {\bf R}_0^{-1} {\bf Z}\right)^{-1} 
{\bf T}^\top
\right).
\label{gloss2}
\end{eqnarray}

The following theorem gives the maximum loss function $\phi_{\cal L}$ for the LSE and MGLSE.

\begin{theorem}
For the neighbourhoods ${\bf R}_{{\bf K}, \alpha}$ defined in (\ref{NBR}) and 
measure ${\cal L}$ being monotonic according to the ordering of positive definiteness, we have
\begin{eqnarray}
&& \hspace{-1.2cm} \mbox{(i)} ~\phi_{\cal L} 
\left( \hat{\mbox{\boldmath $\mu$}}_L, {\bf X} \right) 
=\left\{ \begin{array}{ll}
{\cal L} \left( \frac{1+\alpha}{b^2} ~{\bf X}^\top {\bf R}_0 {\bf X} \right), & \mbox{for}~~ {\bf K}={\bf R}_{j0},  \\
{\cal L} \left( \frac{1}{b^2} ~\left({\bf X}^\top {\bf R}_0 {\bf X}  + \alpha {\bf X}^\top {\bf X} \right)\right),
& \mbox{for}~~ {\bf K}={\bf I}_t,
\end{array} \right.
\label{maxLoss2} \\
&& \hspace{-1.2cm} \mbox{(i)} ~\phi_{\cal L} 
\left( \hat{\mbox{\boldmath $\mu$}}_M, {\bf X} \right) 
=\left\{ \begin{array}{ll}
{\cal L} \left( (1+\alpha)
{\bf T} 
\left({\bf Z}^\top {\bf R}_0^{-1} {\bf Z}\right)^{-1}
{\bf T}^\top
\right),  & \mbox{for}~~ {\bf K}={\bf R}_{j0},  \\
{\cal L} \left(
{\bf T} 
\left({\bf Z}^\top {\bf R}_0^{-1} {\bf Z}\right)^{-1}
{\bf Z}^\top  {\bf C}_0 {\bf Z}
\left({\bf Z}^\top {\bf R}_0^{-1} {\bf Z}\right)^{-1} 
{\bf T}^\top
\right),  & \mbox{for}~~{\bf K}={\bf I}_t,
\end{array} \right.
\label{maxLoss3}
\end{eqnarray}
where ${\bf C}_0= {\bf R}_0^{-1}  + \alpha {\bf R}_0^{-2}$.
\end{theorem}

The proof of Theorem 1 is given in the Appendix. The results in Theorem 1 are very useful, and
we only need to minimize (\ref{maxLoss2}) or (\ref{maxLoss3}) to construct robust designs.
In the next section, we will discuss two algorithms to find robust designs, present representative 
examples, and derive several theoretical results.

\section{Construction of robust designs}

\subsection{Numerical algorithms}

Minimizing (\ref{maxLoss2}) or (\ref{maxLoss3}) over ${\bf X}$ is a combinatorial optimization problem.
When the number of treatments and the number of blocks are small, a complete search method
to find robust designs is feasible.
However, when the number of treatments and/or the number of blocks are big,
it is too expensive to use a complete search method.  In this situation,
there are various algorithms available that can be  applied to construct robust designs.
One of them is a simulated annealing algorithm, which is known in the literature to be effective in searching for optimal 
and robust designs.  
For example, see Elliott, Eccleston and Martin (1999), Fang and Wiens (2000), and
Wilmut and Zhou (2011).

An annealing algorithm  minimizing 
$\phi_{\cal L} \left( \hat{\mbox{\boldmath $\mu$}}, {\bf X} \right)$ 
includes the following main steps.
\begin{description}
\item{Step 1:}  Choose an initial design ${\bf X}$, say ${\bf X}_0$, 
and set initial values of the parameters in the algorithm
such as the cooling temperature and the number of iterations at each temperature amongst others.
Compute the maximum loss function at ${\bf X}_0$ as
$l_0=\phi_{\cal L} 
\left( \hat{\mbox{\boldmath $\mu$}}, {\bf X}_0 \right)$. 

\item{Step 2:}  Use a scheme to generate a new design,  say ${\bf X}_1$,  
which is usually a small change from the current design ${\bf X}_0$. 
Compute the maximum loss function at ${\bf X}_1$ as
$l_1=\phi_{\cal L} 
\left( \hat{\mbox{\boldmath $\mu$}}, {\bf X}_1 \right)$. 

\item{Step 3:}  Use a rule to determine if the new design ${\bf X}_1$ can be accepted.  If it is accepted, then it becomes ${\bf X}_0$.

\item{Step 4:}  Update the cooling temperature.  Use a stopping criterion to see if the designs have converged.  If 
converged,  go to Step 5.  Otherwise, go to Step 2.

\item{Step 5:}  The last design ${\bf X}_0$ is considered to be an approximate optimal design.
\end{description}

The cooling temperature parameter,  $T$, plays an important role in the algorithm, and it has influence on
the speed of convergence of the designs.  
The detailed discussions about setting the initial cooling temperature and how to update it 
can be found in Fang and Wiens (2000) and the references therein.
The acceptance rule is as follows.
If $l_1 \le l_0$, then ${\bf X}_1$ is accepted.
If $l_1 > l_0$, then ${\bf X}_1$ is accepted with a probability $\exp(-(l_1-l_0)/T)$.

At each iteration, a new design needs to be generated, and it is usually obtained by modifying the current
design with a small change.  A good scheme for generating new designs should allow us to access all possible designs 
for ${\bf X}$.  Since we can randomly assign the numbers to the $t$ treatments, without loss of generality we fix the
allocation of treatments in block 1, and only search for optimal allocations in blocks 2 to $b$.
A new design ${\bf X}_1$ is obtained from ${\bf X}_0$ by
randomly choosing a block number from 2 to $b$ and switching two treatment numbers in the selected block.

There are other modifications that can improve the searching. Two small steps are added in our computation.
One is to record the best design, say ${\bf X}_*$,  with the
smallest $\phi_{\cal L}$ during the iterations. 
Notice that ${\bf X}_*$ is updated at each iteration.
At the end, if ${\bf X}_*$ has smaller $\phi_{\cal L}$ value than ${\bf X}_0$, then  ${\bf X}_*$ is considered as 
an approximate robust design. 
Another step is to start with the approximate robust design from the annealing algorithm 
and do an additional steepest descent procedure as in  Elliott, Eccleston and Martin (1999).  
This can be done by running the above annealing algorithm again and 
only accepting the new design when $l_1 \le l_0$.


\subsection{Applications}

We consider a general setting for each block in the following examples.
Suppose each block contains $m \times n$ small plots arranged in a rectangular area as in Table \ref{table3}.
This is common for field experiments, and each small plot receives a treatment.
Assume each block 
has one replicate of $t$ treatments, so $t=mn$.
Let ($k$, $s$)  indicate the position of  a small plot, $k=1, \cdots, m, ~~s=1, \cdots, n$.
In model (\ref{reg1}),  we define the error vector for the $j$th block as
\begin{eqnarray*}
\mbox{\boldmath $\epsilon$}_j=\left(
\epsilon_{1,1}, \cdots, \epsilon_{1,n}, \epsilon_{2,1}, \cdots, \epsilon_{2,n}, 
\cdots,
\epsilon_{m,1}, \cdots, \epsilon_{m,n} \right)^\top, ~~~j=1, \cdots, b.
\end{eqnarray*}

\begin{table}
\caption{Small plots arrangement in each block}
\begin{center}
\begin{tabular}{|c|c|c|c|} \hline
(1,1) & (1,2) & ~~~~~~$\cdots$~~~~~~ & (1,$n$) \\ \hline
(2,1) & (2,2) & $\cdots$ & (2,$n$) \\ \hline
$\vdots$ & $\vdots$ & $\vdots$ & $\vdots$ \\ \hline
($m$,1) & ($m$,2) & $\cdots$ & ($m$,$n$) \\ \hline
\end{tabular}
\end{center}
\label{table3}
\end{table}

If it is not a field experiment but the runs in each
block  are conducted over time, then it  can be viewed as a special case with $n=1$.
Four representative examples are presented next to show the robust designs.
In all the examples we set $\sigma^2=1$ to present the loss function values.

\noindent{\bf Example 1} Construct the robust design for $b=2, ~t=7, ~m=7$, and $n=1$.  The
neighbourhood is ${\bf R}_{{\bf K},\alpha}$ with $\alpha=0.25$ and ${\bf K}={\bf R}_{j0}=\sigma^2{\bf V}_{j0}$, where
${\bf V}_{j0}$ is from the first order NN process with correlation $\rho=0.15$, {\it i.e.},
\begin{eqnarray*}
{\bf V}_{j0}=
\left( \begin{array}{ccccc}
1 & \rho & 0  & \cdots &0 \\
\rho & 1 & \rho & \cdots &0  \\
 &  \ddots & \ddots & \ddots &  \\
0 & \cdots &  \rho & 1  & \rho \\
0 & \cdots & 0  & \rho & 1 
\end{array} \right)_{7 \times 7}. 
\end{eqnarray*}
Using the MGLSE, we minimize 
$\phi_D 
\left( \hat{\mbox{\boldmath $\mu$}}_M, {\bf X} \right) =
\det\left( (1+\alpha)
{\bf T} 
\left({\bf Z}^\top {\bf R}_0^{-1} {\bf Z}\right)^{-1}
{\bf T}^\top
\right)$ to get the D-optimal robust design. 
Notice that the result does not depend on the value of $\sigma^2$ or $\alpha$.
A complete search method is applied, and the results show that the D-optimal robust design
is not unique.  

\begin{figure}[h!t]
\begin{center}
{
\begin{tabular}{|c|}
\hline
1 \\ \hline
2 \\ \hline
3 \\ \hline
4 \\ \hline
5 \\ \hline
6 \\ \hline
7 \\ \hline
\end{tabular}
\hspace{0.85cm}
\begin{tabular}{|c|}
\hline
7 \\ \hline
5 \\ \hline
2 \\ \hline
4 \\ \hline
6 \\ \hline
3 \\ \hline
1 \\ \hline
\end{tabular}
} 
\caption{Robust design for the MGLSE with $t = 7$, $m=7$, $n = 1$, and $b = 2$. 
The numbers, $1, \cdots, 7$, are treatment numbers in the small plots.}
\label{7treat_onecol_arrangement}
\end{center}
\end{figure}

Figure \ref{7treat_onecol_arrangement} presents one robust design with 
$\left(\phi_D\left( \hat{\mbox{\boldmath $\mu$}}_M, {\bf X} \right) \right)^{1/7}=0.60613$. We notice that
the two blocks have different treatment allocations.  In fact, if two treatments are neighbours in block 1,
then they are not neighbours in block 2.

\bigskip

\noindent{\bf Example 2}  Construct the robust design for $b=5, ~t=3, ~m=3$, and $n=1$.
This experiment is discussed in Section 2.3.
The neighbourhood is ${\bf R}_{{\bf K},\alpha}$ with $\alpha=0.2$ and 
${\bf K}={\bf R}_{j0}=\sigma^2{\bf V}_{j0}$, where
${\bf V}_{j0}$ ($3 \times 3$) is from the first order NN process with correlation $\rho=0.20$.
One D-optimal robust design minimizing $\phi_D 
\left( \hat{\mbox{\boldmath $\mu$}}_M, {\bf X} \right)$
is given in Figure \ref{Example21}, obtained from a complete search method.
This design has 
$\left(\phi_D\left( \hat{\mbox{\boldmath $\mu$}}_M, {\bf X} \right) \right)^{1/3}=0.23165$.

\bigskip
\begin{figure}[h!t]
\begin{small}
\begin{center}
{
\begin{tabular}{|c|}
\hline
1 \\ \hline
2 \\ \hline
3 \\ \hline
\end{tabular}
\hspace{0.85cm}
\begin{tabular}{|c|}
\hline
3 \\ \hline
2 \\ \hline
1 \\ \hline
\end{tabular}
\hspace{0.85cm}
\begin{tabular}{|c|}
\hline
2 \\ \hline
1 \\ \hline
3 \\ \hline
\end{tabular}
\hspace{0.85cm}
\begin{tabular}{|c|}
\hline
3 \\ \hline
1 \\ \hline
2 \\ \hline
\end{tabular}
\hspace{0.85cm}
\begin{tabular}{|c|}
\hline
1 \\ \hline
3 \\ \hline
2 \\ \hline
\end{tabular}
} 
\caption{Robust design for the MGLSE with $t = 3$, $m=3$, $n = 1$, and $b = 5$. 
The numbers, $1, 2, 3$, are treatment numbers in the small plots.}
\label{Example21}
\end{center}
\end{small}
\end{figure}

\bigskip
\begin{figure}[h!t]
\begin{small}
\begin{center}
{
\begin{tabular}{|c|}
\hline
1 \\ \hline
2 \\ \hline
3 \\ \hline
\end{tabular}
\hspace{0.85cm}
\begin{tabular}{|c|}
\hline
1 \\ \hline
2 \\ \hline
3 \\ \hline
\end{tabular}
\hspace{0.85cm}
\begin{tabular}{|c|}
\hline
3 \\ \hline
2 \\ \hline
1 \\ \hline
\end{tabular}
\hspace{0.85cm}
\begin{tabular}{|c|}
\hline
1 \\ \hline
2 \\ \hline
3 \\ \hline
\end{tabular}
\hspace{0.85cm}
\begin{tabular}{|c|}
\hline
3 \\ \hline
2 \\ \hline
1 \\ \hline
\end{tabular}
} 
\caption{Robust design for the LSE with $t = 3$, $m=3$, $n = 1$, and $b = 5$. 
The numbers, $1, 2, 3$, are treatment numbers in the small plots.}
\label{Example22}
\end{center}
\end{small}
\end{figure}

One D-optimal robust design minimizing $\phi_D 
\left( \hat{\mbox{\boldmath $\mu$}}_L, {\bf X} \right)$
is given in Figure \ref{Example22}, which gives 
$\left(\phi_D\left( \hat{\mbox{\boldmath $\mu$}}_L, {\bf X} \right) \right)^{1/3}=0.23342$.  
The D-optimal robust designs are not unique.  
We can randomly permute the treatment numbers and block numbers.  We can also change the orientation of  blocks if matrix
${\bf V}_{j0}$ is from a weakly stationary error process.  This implies that the design in Figure \ref{Example22} is the same as 
design $d1$ in Table \ref{table2}, and the design in Figure \ref{Example21} is the same as 
design $d2$ in Table \ref{table2}.  The D-optimal design based on the LSE puts the same treatment in the middle plot of all the 5 blocks,
while the D-optimal design based on the MGLSE distributes the three treatments in the middle plot almost uniformly.
In addition, if ${\bf K}={\bf I}_t$, we get the same D-optimal robust designs in Figures \ref{Example21} and \ref{Example22} for the
MGLSE and LSE, respectively.

\bigskip

\noindent{\bf Example 3}
Construct the robust design for $b=2, ~t=12, ~m=6$, and $n=2$.  
The neighbourhood is ${\bf R}_{{\bf K},\alpha}$ with $\alpha=0.3$ and 
${\bf K}=\sigma^2{\bf I}_{t}$.  Take $\sigma^2=1$.
The correlation matrix  ${\bf V}_{j0}$ ($12 \times 12$) is from the DG with parameter $\lambda$,
{\it i.e.}, the correlation between two small plots at locations $(k_1,s_1)$ and $(k_2,s_2)$ is given by
$\lambda^{|k_1-k_2|+|s_1-s_2|}$.  Robust designs are found using the annealing algorithm and are presented
for two values of $\lambda$ in Figure \ref{Dopt_DG12}.  We have 
$\left(\phi_D\left( \hat{\mbox{\boldmath $\mu$}}_M, {\bf X} \right) \right)^{1/12}=0.64993$ and $0.58963$ 
for $\lambda=0.01$ and  $ 0.3$, respectively.  


\begin{figure}[h!t]
\begin{small}
\begin{center}
{(a)~~
\begin{tabular}{|c|c|}
\hline
1 & 2 \\ \hline
3 & 4 \\ \hline
5 & 6 \\ \hline
7 & 8 \\ \hline
9 & 10 \\ \hline
11 & 12 \\ \hline
\end{tabular}
\hspace{0.25cm}
\begin{tabular}{|c|c|} 
\hline
11 & 2 \\ \hline
6 & 7 \\ \hline
3 & 10 \\ \hline
9 & 4 \\ \hline
8 & 5 \\ \hline
1 & 12 \\ \hline
\end{tabular}
}
\hspace{2cm}
{(b)~~
\begin{tabular}{|c|c|}
\hline
1 & 2 \\ \hline
3 & 4 \\ \hline
5 & 6 \\ \hline
7 & 8 \\ \hline
9 & 10 \\ \hline
11 & 12 \\ \hline
\end{tabular}
\hspace{0.25cm}
\begin{tabular}{|c|c|} 
\hline
6 & 12 \\ \hline
9 & 3 \\ \hline
2 & 8 \\ \hline
5 & 11 \\ \hline
10 & 4 \\ \hline
1 & 7 \\ \hline
\end{tabular}
}
\caption{Robust designs for the MGLSE with $t = 12$, $m=6$, $n = 2$, and $b = 2$ under the DG:  
(a) $\lambda = 0.01$, (b) $\lambda = 0.3$.
}
\label{Dopt_DG12}
\end{center}
\end{small}
\end{figure}

\bigskip

\noindent{\bf Example 4}  Robust designs are constructed for $b=2$, $n=2$ and various values of $t$.
The first order NN correlation structure is used and $\rho=0.2$, and the neighborhood is
${\bf R}_{{\bf K},\alpha}$ with $\alpha=0.3$ and ${\bf K}={\bf R}_{j0}$.
The designs are presented in Figure \ref{Dopt_NN_vart}, and they
minimize $\phi_D 
\left( \hat{\mbox{\boldmath $\mu$}}_M, {\bf X} \right)$.
The minimum loss function values are 
$\left(\phi_D\left( \hat{\mbox{\boldmath $\mu$}}_M, {\bf X} \right) \right)^{1/t}=
0.59115, 0.58950, 0.58823, 0.58734$ and $0.58662$
for $t=10, 12, 14, 16$ and $18$, respectively.
All the designs have the property that the neighbours in block one are not neighbours in block two.

\begin{figure}[h!t]
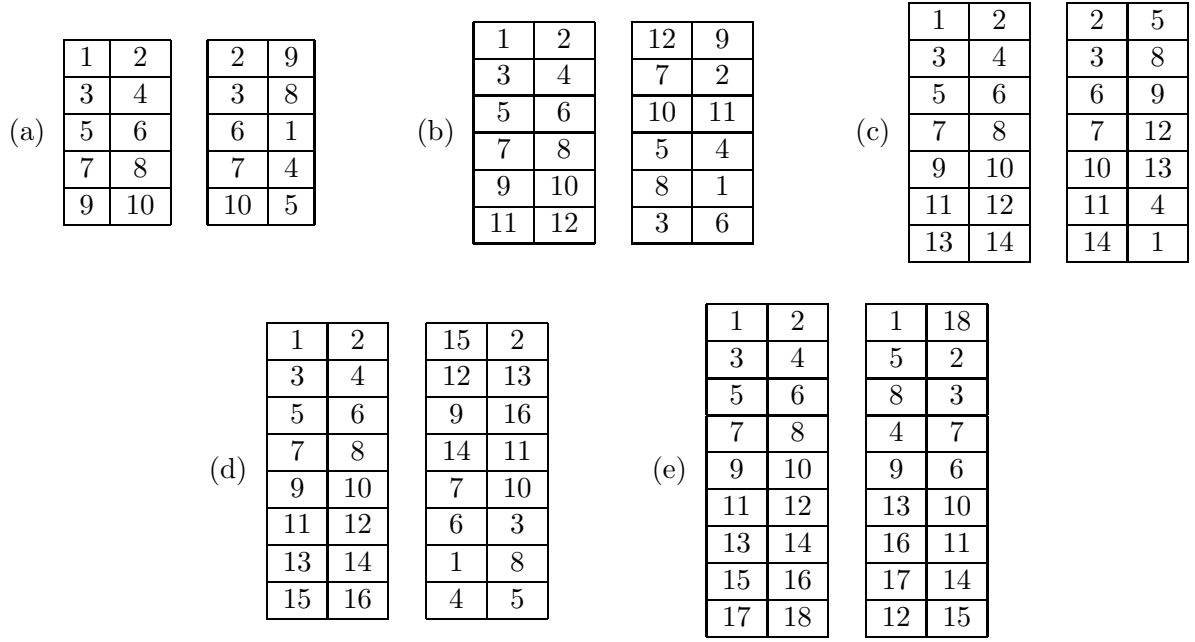

\begin{small}
\begin{center}
{(a)~
\begin{tabular}{|c|c|}
\hline
1 & 2 \\ \hline
3 & 4 \\ \hline
5 & 6 \\ \hline
7 & 8 \\ \hline
9 & 10 \\ \hline
\end{tabular}
\hspace{0.25cm}
\begin{tabular}{|c|c|} 
\hline
2 & 9 \\ \hline
3 & 8 \\ \hline
6 & 1 \\ \hline
7 & 4 \\ \hline
10 & 5 \\ \hline
\end{tabular}
}
\hspace{1.0cm}
{(b)~
\begin{tabular}{|c|c|}
\hline
1 & 2 \\ \hline
3 & 4 \\ \hline
5 & 6 \\ \hline
7 & 8 \\ \hline
9 & 10 \\ \hline
11 & 12 \\ \hline
\end{tabular}
\hspace{0.25cm}
\begin{tabular}{|c|c|} 
\hline
12 & 9 \\ \hline
7 & 2 \\ \hline
10 & 11 \\ \hline
5 & 4 \\ \hline
8 & 1 \\ \hline
3 & 6 \\ \hline
\end{tabular}
}
\hspace{1.0cm}
{(c)~
\begin{tabular}{|c|c|}
\hline
1 & 2 \\ \hline
3 & 4 \\ \hline
5 & 6 \\ \hline
7 & 8 \\ \hline
9 & 10 \\ \hline
11 & 12 \\ \hline
13 & 14 \\ \hline
\end{tabular}
\hspace{0.25cm}
\begin{tabular}{|c|c|} 
\hline
2 & 5 \\ \hline
3 & 8 \\ \hline
6 & 9 \\ \hline
7 & 12 \\ \hline
10 & 13 \\ \hline
11 & 4 \\ \hline
14 & 1 \\ \hline
\end{tabular}
}
\newline
\newline
\newline
{(d)~
\begin{tabular}{|c|c|}
\hline
1 & 2 \\ \hline
3 & 4 \\ \hline
5 & 6 \\ \hline
7 & 8 \\ \hline
9 & 10 \\ \hline
11 & 12 \\ \hline
13 & 14 \\ \hline
15 & 16 \\ \hline
\end{tabular}
\hspace{0.25cm}
\begin{tabular}{|c|c|}  
\hline
15 & 2 \\ \hline
12 & 13 \\ \hline
9 & 16 \\ \hline
14 & 11 \\ \hline
7 & 10 \\ \hline
6 & 3 \\ \hline
1 & 8 \\ \hline
4 & 5 \\ \hline
\end{tabular}
}
\hspace{1.0cm}
{(e)~
\begin{tabular}{|c|c|}
\hline
1 & 2 \\ \hline
3 & 4 \\ \hline
5 & 6 \\ \hline
7 & 8 \\ \hline
9 & 10 \\ \hline
11 & 12 \\ \hline
13 & 14 \\ \hline
15 & 16 \\ \hline
17 & 18 \\ \hline
\end{tabular}
\hspace{0.25cm}
\begin{tabular}{|c|c|} 
\hline
1 & 18 \\ \hline
5 & 2 \\ \hline
8 & 3 \\ \hline
4 & 7 \\ \hline
9 & 6 \\ \hline
13 & 10 \\ \hline
16 & 11 \\ \hline
17 & 14 \\ \hline
12 & 15 \\ \hline
\end{tabular}
}
\caption{Robust designs for the MGLSE for various $t$ under the first order NN: 
(a) $t = 10$, (b) $t = 12$, (c) $t = 14$, (d) $t = 16$, and (e) $t = 18$.}
\label{Dopt_NN_vart}
\end{center}
\end{small}
\end{figure}

\bigskip

\subsection{Theoretical properties}

Analytical solutions for robust designs are hard to obtain in general, but 
we are able to derive several theoretical results for block designs here.


\begin{theorem}
For neighbourhood ${\bf R}_{{\bf R}_{j0}, \alpha}$ with  
${\bf R}_0={\bf R}_{10} \oplus {\bf R}_{10} \oplus \cdots \oplus {\bf R}_{10}$ ,
the design with the same treatment allocation in all the $b$ blocks is a D-optimal robust design, which
minimizes $\phi_D \left( \hat{\mbox{\boldmath $\mu$}}_L, {\bf X} \right)$.
\end{theorem}

The proof of Theorem 2 is in the Appendix.  The result is true for any ${\bf R}_{10}$, $b \ge 2$ and $n \ge 1$.
As indicated in Example 2, 
we can also permute the treatment numbers and block numbers in the robust designs.  In addition,
if ${\bf R}_{10}$ is from a weakly stationary error process, then we can change the orientation of any number of blocks in the
D-optimal robust designs.

\begin{theorem}
For the LSE and neighbourhood ${\bf R}_{{\bf K}, \alpha}$ with ${\bf K}={\bf R}_{j0}$ or ${\bf I}_t$,
any design is an A-optimal robust design, which minimizes $\phi_A \left( \hat{\mbox{\boldmath $\mu$}}_L, {\bf X} \right)$.
\end{theorem}

The proof of Theorem 3 is in the Appendix.  The result is true for any number of blocks and $n \ge 1$. 
The result implies that the trace is not a good measure to differentiate the designs
for the LSE and neighbourhood ${\bf R}_{{\bf K}, \alpha}$.

\begin{theorem}
Consider block designs with $b=2$, $n=1$ and $t >3$. For the MGLSE and neighbourhood ${\bf R}_{{\bf R}_{j0}, \alpha}$ with ${\bf V}_{j0}$
being the DG or DE correlation matrix, 
the D-optimal robust design, which minimizes $\phi_D \left( \hat{\mbox{\boldmath $\mu$}}_M, {\bf X} \right)$, does not have
the same treatment allocation in the two blocks.
\end{theorem}

The proof of Theorem 4 is in the Appendix.  The result shows that the D-optimal robust designs based on the LSE and the MGLSE are different.
In addition, from the proofs of Theorems 2 and 4,
we can see that
$$ \min_{\bf X} \phi_D \left( \hat{\mbox{\boldmath $\mu$}}_M, {\bf X} \right) < 
\frac{(1+\alpha)^t \sigma^{2t}}{2^t} ~\det({\bf V}_{10}) = \min_{\bf X} \phi_D \left( \hat{\mbox{\boldmath $\mu$}}_L, {\bf X} \right).$$
Thus the MGLSE should be applied when there is information about the error correlation.
Theorem 4 is for a specific situation, but we conjecture that the result is true in general. This could be a future research topic.

\subsection{Guidelines for using  robust designs}

{\bf 
Robust designs studied in this paper can be applied to any block experiment in which there is a possibility of correlated errors.
Here is a detailed procedure for practical applications.
\begin{description}
\item{(1)} Specify the block experiment parameters, $t, b, m$ and $n$.

\item{(2)} Use prior information  to propose a
correlation matrix of the errors for each block,  ${\bf V}_{i0}, ~i=1, \cdots, b$. Then
the covariance matrix is ${\bf R}_0=\sigma^2 {\bf V}_{10} \oplus \cdots \oplus \sigma^2{\bf V}_{b0}$.
These correlation matrices may be from the NN, MA, DG, or other error processes, and often we can set ${\bf V}_{10} = \cdots = {\bf V}_{b0}$
if block conditions are similar.

\item{(3)}  Choose the covariance neighbourhood size parameter $\alpha \ge 0$.
If the prior information of the correlation matrix of the errors is very accurate, 
then set $\alpha$ to be very small, say $\alpha=0.10$.
Otherwise, choose a slightly bigger $\alpha$, say $\alpha=0.30$. 

\item{(4)}  Construct the robust design by minimizing 
$\phi_{\cal L} 
\left( \hat{\mbox{\boldmath $\mu$}}_M, {\bf X} \right) $  in (\ref{maxLoss3}).
We can use either ${\bf K}={\bf R}_{j0}=\sigma^2 {\bf V}_{j0}$ or ${\bf K}={\bf I}_t$.

\item{(5)}  Use the robust design to run the experiment and collect data.  After the experiment,
estimate $\mbox{\boldmath $\theta$}$ by $\hat{\mbox{\boldmath $\theta$}}_M$ in (\ref{GLSE2}).

\end{description}

Since we never know the exact covariance matrix of the errors in practice, 
robust designs  perform well in a neighborhood
of the covariance matrix ${\bf R}_0$.  In addition, 
our study indicates that robust designs are not very sensitive to the choices of
${\bf R}_0$ and $\alpha$, from many examples we have constructed.
For instance, in Example 1 the robust design does not depend on the value of $\alpha$, and  the robust design is
highly efficient for a range of $\rho$ values in ${\bf V}_{j0}$.  

We define an efficiency measure to compare a design ${\bf X}_d$ with the robust design ${\bf X}_R$,
\begin{eqnarray*}
\mbox{Eff}(d) = \frac{\phi_{D} 
\left( \hat{\mbox{\boldmath $\mu$}}_M, {\bf X}_R \right) }{\phi_{D} 
\left( \hat{\mbox{\boldmath $\mu$}}_M, {\bf X}_d \right) }.
\end{eqnarray*}
For Example 1, 
we compute the efficiencies for the following three representative designs ${\bf X}_{d}, {\bf X}_{e}$, and ${\bf X}_{f}$.  
The  first block of the three designs
is the same as in ${\bf X}_R$.  The orderings of the second block for the
three designs are as follows,
${\bf X}_{d}$: 7, 6, 5, 4, 3, 2, 1;
${\bf X}_{e}$:  1, 2, 3, 4, 7, 6, 5;
${\bf X}_{f}$:  2, 1, 4, 3, 6, 5, 7.
The efficiencies are given in Table \ref{Table4}.  It is clear that the robust design is more efficient than the three designs
for a range of $\rho$ values.
}

\begin{table}
\caption{Efficiencies for three designs compared with the robust design in Example 1.}
\begin{center}
\begin{tabular}{l|rrrrr} \hline
$\rho$   &  0.10   &  0.15   & 0.20  &  0.25  & 0.30  \\ \hline
\mbox{Eff}(d)  & 0.969 & 0.927 & 0.868 & 0.790 & 0.692 \\
\mbox{Eff}(e)  & 0.973 & 0.937 & 0.885 & 0.816 & 0.728  \\
\mbox{Eff}(f)  & 0.983 & 0.961 & 0.928 & 0.882 & 0.821 \\ \hline
\end{tabular}
\label{Table4}
\end{center}
\end{table}

\section{Conclusion}

We have investigated robust designs for experiments running in $b$ blocks, 
where a complete replicate of  $t$ treatments is run in each block.
These designs are robust against possible misspecification of the covariance matrix of the errors 
within each block.  We used a neighbourhood
to model the  unknown covariance matrix of the errors instead of specifying it exactly.
Robust designs are defined using a minimax approach, {\it i.e.}, 
 minimizing the maximum loss of $Cov \left( \hat{\mbox{\boldmath $\mu$}} \right)$,
where the estimator can be the LSE or the MGLSE. 
Several interesting theoretical results and examples have been obtained and presented.
In particular, the robust designs based on the LSE and MGLSE are quite different.
The results in this paper indicate that when there is information about the correlation of the errors, the
MGLSE should be used to construct robust designs and to estimate the treatment means.

In this paper we have focused on the block designs with one replicate within each block, 
and a measure of $Cov \left( \hat{\mbox{\boldmath $\mu$}} \right)$ is minimized.
However, the methodology can be easily  extended  to 
situations where
\begin{description}
\item (a) there are two or more replicates in each block,

\item (b) we want to minimize $Cov \left( {\bf C} \hat{\mbox{\boldmath $\mu$}} \right)$, where ${\bf C}$ ($v \times t$)
is a constant contrast matrix, with $v \le t$.
\end{description}

For (a), we only need to make some dimensional changes in matrices ${\bf X}$, ${\bf U}$ and ${\bf V}$.
For (b), notice that $Cov \left( {\bf C} \hat{\mbox{\boldmath $\mu$}} \right) 
= {\bf C}~ Cov \left( \hat{\mbox{\boldmath $\mu$}} \right) {\bf C}^\top$. But interesting results may be derived
for various contrast matrices.
If there is a control in the $t$ treatments, say treatment 1, then it is natural to compare each treatment with the control and
the contrast matrix can be defined as
\begin{eqnarray*}
{\bf C} = \left( \begin{array}{ccccc}
1 & $-1$ & 0 & \cdots & 0 \\
1 & 0 & $-1$ & \cdots & 0 \\
\vdots & &  & \ddots & \\
1 & 0 & 0 & \cdots &$-1$  
\end{array} \right)_{(t-1) \times t}.
\end{eqnarray*}
We can also define other contrast matrices to study  linear combinations of the $t$ treatment means.

\bigskip

\section*{Appendix: Proofs} 

\noindent{\bf Proof of Theorem 1:}  From (\ref{gloss1}), we have
\begin{eqnarray*}
g_{\cal L} \left( \hat{\mbox{\boldmath $\mu$}}_L, {\bf X},  {\bf R} \right) = 
{\cal L} \left( \frac{1}{b^2} ~{\bf X}^\top {\bf R} {\bf X} \right),
\end{eqnarray*}
then from the definition of ${\bf R}_{{\bf K}, \alpha}$ in (\ref{NBR}), we get
\begin{eqnarray*}
&&{\bf R} \le {\bf R}_0 + \alpha {\bf K}_0, ~~~\mbox{for all}~~{\bf R} \in {\bf R}_{{\bf K}, \alpha}, ~~~\mbox{and}\\
&&{\bf X}^\top {\bf R} {\bf X} \le {\bf X}^\top  ({\bf R}_0 + \alpha {\bf K}_0) {\bf X},
\end{eqnarray*}
where ${\bf K}_0={\bf K} \oplus  {\bf K} \oplus + \cdots \oplus  {\bf K}$, with the same dimensions as matrix ${\bf R}_0$.
Since measure ${\cal L}$ is monotonic according to the ordering of positive definiteness,
it is clear that
\begin{eqnarray*}
g_{\cal L} \left( \hat{\mbox{\boldmath $\mu$}}_L, {\bf X},  {\bf R} \right) \le
{\cal L} \left( \frac{1}{b^2} ~{\bf X}^\top ({\bf R}_0 + \alpha {\bf K}_0) {\bf X} \right),
~~~\mbox{for all}~~{\bf R} \in {\bf R}_{{\bf K}, \alpha}.
\end{eqnarray*} 
Thus, from (\ref{maxLoss1}),
\begin{eqnarray*}
\phi_{\cal L} 
\left( \hat{\mbox{\boldmath $\mu$}}_L, {\bf X} \right) =
{\cal L} \left( \frac{1}{b^2} ~{\bf X}^\top ({\bf R}_0 + \alpha {\bf K}_0) {\bf X} \right).
\end{eqnarray*}
Putting ${\bf K}={\bf R}_{j0}$ and $ {\bf K}={\bf I}_t$ in the above equation gives the results in 
(\ref{maxLoss2}).
For the MGLSE, the covariance matrix is in (\ref{COVM}) and the loss function is in (\ref{gloss2}).
By a similar proof to the proof for the LSE above, we can get the result in 
(\ref{maxLoss3}).

\bigskip

\noindent{\bf Proof of Theorem 2:}  From (\ref{maxLoss2}), we have
\begin{eqnarray*}
\phi_D
\left( \hat{\mbox{\boldmath $\mu$}}_L, {\bf X} \right) 
=\det \left( \frac{1+\alpha}{b^2} ~{\bf X}^\top {\bf R}_0 {\bf X} \right)=
\frac{(1+\alpha)^t}{b^{2t}} ~ \det \left( {\bf X}^\top {\bf R}_0 {\bf X} \right).
\end{eqnarray*}
Notice that ${\bf X}^\top = ({\bf X}_1^\top, {\bf X}_2^\top, \cdots, {\bf X}_b^\top)$ and
${\bf R}_0={\bf R}_{10} \oplus {\bf R}_{10} \oplus \cdots \oplus {\bf R}_{10}$, which gives 
\begin{eqnarray}
{\bf X}^\top {\bf R}_0 {\bf X}  = {\bf X}_1^\top {\bf R}_{10} {\bf X}_1 + {\bf X}_2^\top {\bf R}_{10} {\bf X}_2
+ \cdots + {\bf X}_b^\top {\bf R}_{10} {\bf X}_b.
\label{TH21}
\end{eqnarray}
Since the treatment labels are randomly assigned, without loss of generality we can number the treatments in block 1 such 
that ${\bf X}_1={\bf I}_t$.  In addition we can write matrix ${\bf X}_j={\bf X}_1 {\bf P}_j={\bf P}_j$, where
${\bf P}_j$ is a ($t \times t$) permutation matrix, $j=2, \cdots, b$.  It is obvious that ${\bf P}_j^\top ={\bf P}_j^{-1}$.
Then from (\ref{TH21}),  we get
\begin{eqnarray}
\det \left( {\bf X}^\top {\bf R}_0 {\bf X} \right) =
\det \left( {\bf R}_{10} + {\bf P}_2^\top {\bf R}_{10} {\bf P}_2 + \cdots + {\bf P}_b^\top {\bf R}_{10} {\bf P}_b\right).
\label{Th22}
\end{eqnarray}
Define ${\bf A}_1={\bf R}_{10}$, ${\bf A}_j={\bf P}_j^\top {\bf R}_{10} {\bf P}_j, ~j=2, \cdots, b$, and
${\bf A}=\frac{1}{b} ({\bf A}_1 + {\bf A}_2 +\cdots +{\bf A}_b)$. It is obvious that 
$\det({\bf A}_j)=\det({\bf A}_1)$, since $\det({\bf P}_j)=1$.
Using Minkowski's inequality in Horn and Johnson (1985, page 482), we can show that 
$\det({\bf A}) \ge \det({\bf A}_1)$,  where the equality holds  if ${\bf A}_1=  {\bf A}_2 = \cdots = {\bf A}_b$.
Thus,  $\det \left( {\bf X}^\top {\bf R}_0 {\bf X} \right)$ in (\ref{Th22}) is minimized when 
${\bf A}_1=  {\bf A}_2 = \cdots = {\bf A}_b$.
This implies that the design with the same treatment allocation in all the $b$ blocks is a 
D-optimal robust design, which minimizes $\phi_D \left( \hat{\mbox{\boldmath $\mu$}}_L, {\bf X} \right)$.

\bigskip

\noindent{\bf Proof of Theorem 3:}  We prove the result for ${\bf K}={\bf R}_{j0}$ here.  The 
result for ${\bf K}={\bf I}_t$ can be proved similarly.  
Notice that for any design, ${\bf X}_j^\top {\bf X}_j = {\bf X}_j {\bf X}_j^\top = {\bf I}_t$, for $j=1, \cdots, b$.
From (\ref{maxLoss2}), we have
\begin{eqnarray*}
\phi_A
\left( \hat{\mbox{\boldmath $\mu$}}_L, {\bf X} \right) 
&=& \mbox{trace} \left( \frac{1+\alpha}{b^2} ~{\bf X}^\top {\bf R}_0 {\bf X} \right) \\
&=& \frac{1+\alpha}{b^2} ~\mbox{trace} \left( \sum_{j=1}^b  {\bf X}_j^\top {\bf R}_{j0} {\bf X}_j \right) \\ 
&=& \frac{1+\alpha}{b^2} ~\sum_{j=1}^b  \mbox{trace} \left( {\bf X}_j^\top {\bf R}_{j0} {\bf X}_j    \right) \\
&=& \frac{1+\alpha}{b^2} ~\sum_{j=1}^b  \mbox{trace} \left( {\bf R}_{j0} {\bf X}_j  {\bf X}_j^\top \right)  \\
&=& \frac{1+\alpha}{b^2} ~\sum_{j=1}^b  \mbox{trace} \left( {\bf R}_{j0}\right), 
\end{eqnarray*}
which does not depend on design ${\bf X}$.
Therefore any design is an A-optimal robust design.

\bigskip

\noindent{\bf Proof of Theorem 4:}  For $n=1$, the DG and DE correlation matrix have the same form, which is
given by, for $\lambda \in (0, 1)$,
\begin{eqnarray*}
{\bf V}_{j0}=\left( \begin{array}{ccccc}
1 & \lambda & \lambda^2 & \cdots & \lambda^{t-1} \\
\lambda & 1 & \lambda   & \cdots & \lambda^{t-2} \\
\lambda^2 & \lambda & 1 &  \cdots & \lambda^{t-3} \\
\vdots  & \vdots &  & \ddots & \vdots \\
\lambda^{t-1} & \lambda^{t-2} & \lambda^{t-3} & \cdots & 1 
\end{array}
\right)_{t \times t},  ~~j=1, 2,
\end{eqnarray*}
and it is easy to verify that its inverse matrix is
\begin{eqnarray*}
{\bf V}_{j0}^{-1}= \frac{1}{1-\lambda^2} ~\left( \begin{array}{ccccc}
1 & -\lambda & 0 & \cdots & 0 \\
-\lambda & 1+ \lambda^2 & -\lambda   & \cdots & 0 \\
 & \ddots & \ddots & \ddots &  \\
0 & \cdots &  -\lambda & 1+ \lambda^2 & -\lambda \\
0 &  \cdots & 0 & -\lambda & 1 
\end{array}
\right).
\end{eqnarray*}
For $b=2$, ${\bf R}_0=\sigma^2  {\bf V}_{10} \oplus \sigma^2  {\bf V}_{10}$, 
${\bf T}=({\bf I}_t, {\bf 0})_{t \times (t+1)}$,
and ${\bf Z}=\left( \begin{array}{cr}
{\bf X}_1 & {\bf 1}_t \\
{\bf X}_2 & -{\bf 1}_t 
\end{array} \right)$, where ${\bf 1}_t$ ($t \times 1$) is a vector of ones. 
As in the proof of Theorem 2, let ${\bf X}_1={\bf I}_t$ and ${\bf X}_2={\bf X}_1 {\bf P}_t$,
where ${\bf P}_t$ is a permutation matrix. Then
straightforward calculation gives
\begin{eqnarray*}
{\bf Z}^\top {\bf R}_0^{-1} {\bf Z}= \frac{1}{\sigma^2} ~\left(
\begin{array}{cc}
{\bf V}_{10}^{-1} + {\bf P}_t^\top {\bf V}_{10}^{-1} {\bf P}_t & {\bf V}_{10}^{-1} {\bf 1}_t - {\bf P}_t^\top {\bf V}_{10}^{-1} {\bf 1}_t \\
{\bf 1}_t^\top {\bf V}_{10}^{-1} - {\bf 1}_t^\top {\bf V}_{10}^{-1} {\bf P}_t & 2 ~{\bf 1}_t^\top {\bf V}_{10}^{-1} {\bf 1}_t 
\end{array} \right).
\end{eqnarray*}
Let $c_0= 2 ~{\bf 1}_t^\top {\bf V}_{10}^{-1} {\bf 1}_t = \frac{2}{1-\lambda^2} \left(t - 2(t-1) \lambda + (t-2)\lambda^2 \right)$.
It is clear that $c_0 > 0$.
Now  we have
\begin{small}
\begin{eqnarray*}
{\bf T} 
\left({\bf Z}^\top {\bf R}_0^{-1} {\bf Z}\right)^{-1}
{\bf T}^\top = \sigma^2 
\left( ({\bf V}_{10}^{-1} + {\bf P}_t^\top {\bf V}_{10}^{-1} {\bf P}_t) - \frac{1}{c_0}~
({\bf V}_{10}^{-1} {\bf 1}_t - {\bf P}_t^\top {\bf V}_{10}^{-1} {\bf 1}_t) 
({\bf 1}_t^\top {\bf V}_{10}^{-1} - {\bf 1}_t^\top {\bf V}_{10}^{-1} {\bf P}_t )
\right)^{-1}.
\end{eqnarray*}
\end{small}
If ${\bf P}_t={\bf I}_t$, then 
\begin{eqnarray}
{\bf T} 
\left({\bf Z}^\top {\bf R}_0^{-1} {\bf Z}\right)^{-1}
{\bf T}^\top = \frac{\sigma^2}{2} ~{\bf V}_{10},
\label{SameB1}
\end{eqnarray}
and from (\ref{maxLoss3}), we have
\begin{eqnarray}
\phi_D
\left( \hat{\mbox{\boldmath $\mu$}}_M, {\bf X} \right) 
=\det \left( (1+\alpha)
{\bf T} 
\left({\bf Z}^\top {\bf R}_0^{-1} {\bf Z}\right)^{-1}
{\bf T}^\top
\right)=\frac{(1+\alpha)^t \sigma^{2t}}{2^t} ~\det({\bf V}_{10}).
\label{lossSameB1}
\end{eqnarray}
If ${\bf P}_t \neq {\bf I}_t$ but ${\bf P}_t=1 \oplus {\bf P}_{t-2} \oplus 1$, where ${\bf P}_{t-2}$ is also a permutation matrix,
then it is easy to verify that
${\bf P}_t^\top {\bf V}_{10}^{-1} {\bf 1}_t={\bf V}_{10}^{-1} {\bf 1}_t$. Thus
\begin{eqnarray}
{\bf T} 
\left({\bf Z}^\top {\bf R}_0^{-1} {\bf Z}\right)^{-1}
{\bf T}^\top =\sigma^2  
\left( {\bf V}_{10}^{-1} + {\bf P}_t^\top {\bf V}_{10}^{-1} {\bf P}_t \right)^{-1}.
\label{DiffB1}
\end{eqnarray}
Using Minkowski's inequality in Horn and Johnson (1985, page 482), we can show that 
\begin{eqnarray}
\phi_D
\left( \hat{\mbox{\boldmath $\mu$}}_M, {\bf X} \right) 
&= &\det \left( (1+\alpha)
{\bf T} 
\left({\bf Z}^\top {\bf R}_0^{-1} {\bf Z}\right)^{-1}
{\bf T}^\top
\right)  \nonumber \\
&=& \frac{(1+\alpha)^t \sigma^{2t}}{\det \left( {\bf V}_{10}^{-1} + {\bf P}_t^\top {\bf V}_{10}^{-1} {\bf P}_t \right)}, ~~\mbox{from~}
(\ref{DiffB1})  \nonumber \\
& < & \frac{(1+\alpha)^t \sigma^{2t}}{ 2 ^t ~\det \left( {\bf V}_{10}^{-1} \right) } \nonumber \\
&=& \frac{(1+\alpha)^t \sigma^{2t}}{2^t} ~\det({\bf V}_{10}),
\nonumber
\end{eqnarray}
which is the value of $\phi_D
\left( \hat{\mbox{\boldmath $\mu$}}_M, {\bf X} \right)$ in (\ref{lossSameB1}).  This implies that 
$\phi_D
\left( \hat{\mbox{\boldmath $\mu$}}_M, {\bf X} \right)$ is not minimized by ${\bf P}_t={\bf I}_t$.
This completes the proof.

\bigskip

\section*{Acknowledgements}

This research was partially supported by Discovery Grants from the Natural Sciences and Engineering 
Research Council of Canada. The authors are grateful to the Editor and referee for their helpful comments and suggestions.

\vspace{0.2in}
\renewcommand{\baselinestretch}{1.0}

\noindent {\Large \bf References}
{\small

\begin{description}

\item Bickel, P.J. and Herzberg, A.M. (1979).  Robustness of design against autocorrelation in time I:
Asymptotic theory, optimality for location and linear regression.
{\em Annals of Statistics},  7, 77-95.

\item Bickel, P.J., Herzberg, A.M.,  and Schilling, M.F. (1981). Robustness of design against autocorrelation in time II:
Optimality, theoretical and numerical results for the first-order autoregressive process.
{\em Journal of the American Statistical Association}, 76, 870-877.

\item Box, G.E.P. and Draper, N.R. (1959).  A basis for the selection
of a response surface design.  
{\em Journal of the American Statistical Association}, 54, 622-654.

\item Elliott, L.J., Eccleston, J.A. and Martin, R.J. (1999).
An algorithm for the design of factorial experiments when the data are correlated.
{\em Statistics and Computing}, 9, 195-201.

\item Fang, Z. and Wiens, D.P. (2000).
Integer-valued, minimax robust designs for estimation and extrapolation in heteroscedastic,
approximately linear models.
{\it Journal of the American Statistical Association}, 95, 807-818.

\item Fedorov, V. (2010).
{\it Optimal Experimental Design}.  Wiley, New York.

\item Herzberg, A.M. (1982).
The design of experiments for correlated error structures: Layout and robustness.
{\em Canadian Journal of Statistics}, 10, 133-138.

\item Horn, R. and Johnson, C. (1985).
{\it Matrix Analysis}.  Cambridge University Press, Cambridge.

\item Huber, P.J. (1975).
Robustness and Designs. {\it In A Survey of  Statistical Designs and
Linear Models: Proceedings of an International  Symposium on Statistical
Designs and Linear Models, Colorado State University,} Fort Collins,
March 19-23, 1973, North Holland, Amsterdam, pp. 287-303.

\item Mann, R.K. (2011).
{\it Robust Designs for Field Experiments with Blocks}. 
MSc Thesis, University of Victoria, Victoria, BC, Canada.

\item Martin, R.J. (1982).
Some aspects of experimental design and analysis when errors are correlated.
{\em Biometrika}, 69, 597-612.

\item Martin, R.J. (1986).
On the design of experiments under spatial correlation.
{\em Biometrika}, 73, 247-277 (Correction  75, 396, 1988).

\item Montgomery, D.C. (2012).
{\it Design and Analysis of Experiments}, eighth edition.  Wiley, New York.

\item Ou, B. and Zhou, J. (2009).
Minimax robust designs for field experiments.
{\em Metrika},  69, 45-54.

\item Pukelsheim, F. (1993).  
{\em Optimal Design of Experiments}. Wiley, New York.


%
\item Shi, P., Ye, J. and Zhou, J. (2007).
Discrete minimax designs for regression models with autocorrelated
MA errors.
{\it Journal of Statistical Planning and Inference}, 137, 2721-2731.

\item Wiens, D.P. (1992).
Minimax designs for approximately linear  regression.
{\it Journal of Statistical Planning and Inference,}  31, 353-371.

%
\item Wiens, D.P. and Zhou, J. (1997). 
Robust designs based on the infinitesimal approach.
{\em Journal of the American Statistical Association}, 92, 1503-1511.

\item Wiens, D.P. and Zhou, J. (1999).
Minimax designs for approximately linear models with AR(1) errors.
{\it Canadian Journal of Statistics}, 27, 781-794.

\item Wiens, D.P. and Zhou, J. (2008).
Robust estimators and designs for field experiments.
{\it Journal of Statistical Planning and Inference,}  138, 93-104.

\item Williams, R.M. (1952).  Experimental designs for serially correlated observations.
{\em Biometrika},  39, 151-167.

\item Wilmut, M. and Zhou, J. (2011).
D-optimal minimax design criterion for two-level fractional factorial designs.
{\it Journal of Statistical Planning and Inference,}  141, 576-587.

\item Zhou, J. (2001). Integer-valued, minimax robust designs for 
approximately linear models with correlated errors.
{\em Communications in Statistics: Theory and Methods}, 30, 21-39.

\end{description}             

}


\end{document}